\documentclass[runningheads]{llncs}
\usepackage[T1]{fontenc}
\usepackage{graphicx}
\usepackage{booktabs}
\usepackage{hyperref}
\usepackage[misc]{ifsym}
\newcommand{\corr}{(\Letter)}
\usepackage{amsmath}
\usepackage{amsfonts}
\usepackage{bbold}
\usepackage{subfigure}
\usepackage{placeins}
\usepackage[dvipsnames]{xcolor}
\usepackage{mwe}

\begin{document}

\title{Late fusion ensembles for speech recognition on diverse input audio representations}
\titlerunning{Speech to text conversion using ensembles}

\author{Marin Jezidžić\inst{1} \and
Matej Mihelčić\inst{2},\inst{3} \corr \orcidID{0000-0002-1023-8413}}

\authorrunning{Jezidžić and Mihelčić}

\institute{Project 3 Mobility, Zagreb, Croatia \email{marin.jezidzic@p3m.com}
\and
Faculty of Science, University of Zagreb, Zagreb, Croatia
\email{matmih@math.hr}
\and 
Ruđer Bošković Institute, Zagreb, Croatia
\email{matej.mihelcic@irb.hr}
}




\maketitle              

\begin{abstract}
We explore diverse representations of speech audio, and their effect on a performance of 
late fusion ensemble of E-Branchformer models, applied to Automatic Speech Recognition (ASR) task. Although it is generally known that ensemble methods often improve the performance of the system even for speech recognition, it is very interesting to explore how ensembles of complex state-of-the-art models, such as medium-sized and large E-Branchformers, cope in this setting when their base models are trained on diverse representations of the input speech audio. The results are evaluated on four widely-used benchmark datasets: \textit{Librispeech, Aishell, Gigaspeech}, \textit{TEDLIUMv2} and show that improvements of $1\% - 14\%$ can still be achieved over the state-of-the-art models trained using comparable techniques on these datasets. A noteworthy observation is that such ensemble offers improvements even with the use of language models, although the gap is closing.

\end{abstract}


\section{Introduction}
Despite large volume of work on automatic speech recognition (ASR), an optimal solution for transcribing human speech remains elusive. Among various techniques contributing to the advancements, ensemble methods have emerged as a potent means to enhance model performance by leveraging the strengths of diverse feature or model representations \cite{zhao2014ensemble}. Some techniques \cite{blstm_fusion,MFS,9689650} utilize fusion to tackle ASR. Although there are only a few recent works utilizing ensemble methods for ASR,
 fusion methods with various representations of speech signal are explored in similar problems such as Speaker recognition \cite{Farrell1995}, Emotion recognition \cite{JOTHIMANI2022112512}, Musical instrument recognition \cite{HanETAL} etc. Outside of signal processing, fusion methods are, for example, heavily used in action recognition \cite{sun2023unified} and object detection \cite{s23229162}. As shown in \cite{MFS}, the late fusion approach has the potential to work in an ASR scenario. Late fusion provides flexibility in integrating and optimizing various model outputs, enabling the construction of highly adaptable and efficient ASR systems with inherently parallelizable base model training and inference.
 

Various acoustic representations for ASR \cite{6639051,articleMODGD,845444} were studied concurrently with model development. None of them were tested using current state-of-the-art models. We utilize these representations, as the frontends for our base models \cite{kim2022ebranchformer}. 
When observing ensemble methods, it is of paramount importance to mention the diversity. "If we have a perfect classifier which makes no errors, then we do not need an
ensemble. If however, the classifier does make errors, then we seek to complement
it with another classifier which makes errors on different objects. The diversity of
the classifier outputs is therefore a vital requirement for the success of the ensemble." (Kuncheva, 2014, p. 247) \cite{kunchevaETAL}.

In this work, we utilize several audio signal representations, test their utility for training the state-of-the-art E-Branchformer model for ASR and introduce the generalized decoding method for multiple sound representations making use of multiple scorers per model. This is achieved through the stepwise combination of their predictions. We explore two main questions: can some alternative audio signal representation help the E-Branchformer model achieve better performance than using MEL features? Can ensembles of E-Branchformers trained on diverse audio signal representations improve the performance of SOTA approaches?  The important observation is that the E-Branchformer models, and majority of SOTA ASR models, are very powerful and expressive. Despite this, we demonstrate significant improvements when using ensembles of E-Branchformers, trained on diverse audio signal representations, over the SOTA methodologies in recognition accuracy and occasional improvements of E-Branchformers when trained on the MODGD features instead of MEL. Performed experiments are designed to evaluate the efficacy in various speech recognition scenarios, including noisy environments and speaker variability. 







\section{Related work}
Since the invention of transformer architecture by Vaswani et.al \cite{NIPS2017_3f5ee243}, first used for the task of language translation, and its adaptation for the task of speech recognition \cite{8462506}, there have been many modifications of the base architecture, the learning rate scheduler \cite{Subramanian2023LRS}, various positional information encoding types \cite{pos_enc_overv}, augmentations \cite{inproceedings}, pseudolabelling \cite{pseudolabel}, etc. All these techniques allowed further increase in performance of the system.  Gulati et al. \cite{gulati2020conformer} presented the Conformer architecture which combines the ideas from the Macaron net architecture with relative positional information method \cite{relpos}. The newly developed approach achieved the state of the art results on the Librispeech benchmark dataset \cite{7178964}. The ContextNet \cite{han20_interspeech}, published at a similar time, was not based on the transformer architecture but on the combination of CNN and RNN architectures, showing it is still a viable research option. After these inventions, many new architectures were proposed, one of the more notable are Branchformers \cite{peng2022branchformer} where the encoder is composed of branches, i.e. one branch captures long-range dependencies while the other captures local relationships. The proposed merging method has shown great promise by outperforming Conformer architecture on several benchmarks. This led to research focusing on merging options. The E-Branchformer \cite{kim2022ebranchformer} followed a few months after the Branchformer was presented. E-branchformer holds the best performance up to date with 3.71\% WER on the test\_other subset of \textit{Librispeech} under shallow fusion with LM. On the other hand, newly introduced Zipformer architecture \cite{yao2024zipformer} outperforms it in multiple benchmark datasets such as WenetSpeech \cite{zhang2022wenetspeech} and Aishell \cite{aishell_2017} and without fusing E-branchformer with LM in \textit{Librispeech}. Memory efficient solutions were developed as well \cite{NEURIPS2022_3ccf6da3,ge-etal-2022-edgeformer}, where the compression modules are employed as the replacement for attention, dense and convolutional modules. Edgeformer also explores the usage of LORA (Low order rank adaptation) method \cite{hu2022lora}, which is popular nowadays for compression of LLM models.
Work exploring fusion methods for ASR \cite{blstm_fusion} shows that the combination of Mel spectrograms with varied window size can be combined within the multi-stream HMM fusion architecture. For encoding the input sequence, the authors utilized a simple convolutional network. The PIPO-BLSTM is used for the iterative posterior refinement and decoding of the sequence is performed using weighted finite state transducer (WFST) \cite{MOHRI200269}. In \cite{9689650}, the spectrogram representation was fused with an enhanced version for improved ASR performance. A three layer BLSTM was used to remove the noise as an speech enhancement module. Upon the fusion of spectrograms, the Mel filterbanks were applied to compress the representation and such used as an input to speech recognition model. They make use of the joint training and show it outperforms the case when using separate, pre-trained modules. Finally, work in \cite{MFS} explores the fusion methods between magnitude-based features and phase-based features. They have shown that the fusion techniques, in general, successfully utilize the complementarity between these features. 

\section{Preliminaries}
We evaluate the performance of various features not previously tested on modern speech recognition models. While Mel spectrograms (MEL) and MFCCs are well-known and consistently used in such tasks \cite{10174566}, our focus is on exploring alternative feature types. Gammaton filterbanks (GAMMA) have asymmetric shape, more adapt to how human ear processes sounds. They usually perform better than Mel filterbanks in noisy conditions \cite{YIN2011707}. GAMMA extraction is computationally more expensive than Mel filterbanks. The Constant - Q transform (CQT) utilizes filters that are logarithmically spaced in frequency, providing higher frequency resolution at lower frequencies. It captures harmonic structures better than linear frequency spacing. It excels in low-frequency resolution while maintaining a consistent quality factor, hence Constant-Q, allows for better pitch tracking and harmonic analysis \cite{articleCQT}. \textbf{Modified Group Delay spectrogram} (MODGD) is defined as the negative derivative of the phase of the Fourier transform. It combines phase and magnitude information, enhancing the representation of important spectral features while being robust to noise. MODGD provides good temporal resolution, but not the same level of frequency detail as GAMMA and CQT. Computation of MODGD features involves both magnitude and phase information from the Fourier transform, but is less computationally expensive than CQT and Gamma \cite{articleMODGD}. \textbf{Symlets} (SYMLET) \cite{wavSymlet} represent the family of wavelets utilized in Wavelet transform and allow finer time-frequency resolution across high and low-frequency components. Symlets are similar to Daubechies wavelets but add symmetry for improved phase preservation, balancing time and frequency localization for detailed signal analysis. Computational complexity is usually reduced by employing entropy to find sufficient trees to reconstruct the signal.

\section{Generalization of the joint CTC-Attention Decoding}
\label{sec:method}

In order to make use of the multiple models trained on the diverse representations, we need to find a method of combining their predictions. Each model possesses 2 scorers, the causal attention decoder, and CTC scorer. Since we are dealing with sequence-to-sequence task, we should combine their predictions stepwise. In our work we employ the simple generalization of beam search decoding method proposed by Watanabe et al. \cite{decoding}. The aforementioned approach relies on pre-beam computation based on the output scores of attention decoder to preserve the computational efficiency since the CTC prefix search attains high compute complexity when observing all possible outputs. The proposed workaround for this is combining the scores from multiple attention decoders to obtain more robust pre-beam keys. Since the computation of decoder steps for multiple models can be parallelized, this doesn't add much to the compute complexity. After obtaining the scores from partial scorers, i.e. CTC \& LM, we combine them to obtain the final scores. Following the notation from \cite{decoding}, we denote encoder output as $\textbb{h}$, probability distribution generated by the attention decoder with $p^{att}$ and the probability distribution generated by encoder CTC with $p^{ctc}$. $y=(y_1,y_2,...,y_n)$ denotes output sequence up to point $n \geq 1$ and $\lambda \geq 0$ the tuning parameter. At each step of decoding process, we observe:
\begin{equation}\label{scoring_one_model}
\text{Score} = \lambda\log p^{ctc}(\hat{y}_n|\hat{y}_{1:n-1},\textbb{h}) + (1-\lambda)\log p^{att}(\hat{y}_n|\hat{y}_{1:n-1},\textbb{h})
\end{equation}

In the late fusion setting,  we obtain Score from multiple representations. To compute the Combined score, we use the basic weighted combination.  
\begin{equation}\label{combiner}
 \text{Combined score} = \sum_{i=1}^{\text{num\_repr} }\sum_{j=1}^{2}\alpha_{ij}\text{Score}_{ij}
 \end{equation}

\begin{figure}[!t]
\centering
\includegraphics[scale=0.3,width=9cm]{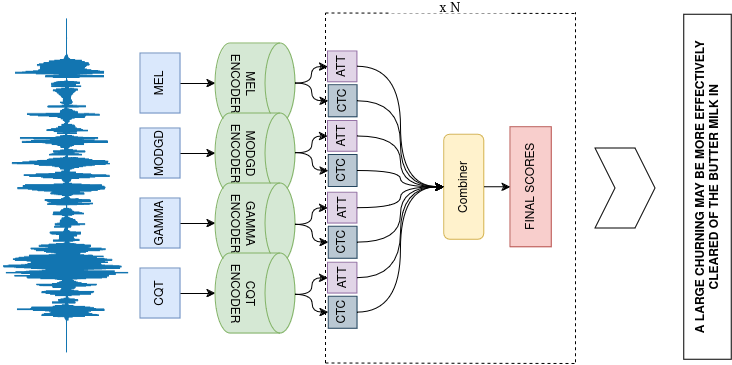}
\caption{Visualization of  decoding process (CTC/AED) ensemble} \label{fig2}
\end{figure}

Where $\alpha_{ij}\geq 0$ and $\sum_{i}\sum_{j} \alpha_{ij} = 1$. The computation of one step is parallelizable with respect to the full scorers and partial scorers. The integration of language model follows \cite{decoding}. The decoding process for single utterance can be observed in figure 
\ref{fig2} where $N$ denotes total number of tokens.

 Approach for late fusion proposed in \cite{MFS} is a trivial case of our approach where the CTC weight is 0. Watanabe et al. \cite{decoding} has shown that the Hybrid approach yields better results overall. The approach in \cite{blstm_fusion} uses the outdated architecture which is not easily adaptable towards the more modern models.

\section{Experiments}
\label{sec:exp}
The goal of our investigation is to determine the benefits of employing a late fusion ensemble approach. Thus, we do not compare our results with methods comprising usage of additional datasets such as Whisper from OpenAI \cite{radford2022robust} and Wav2Vec \cite{baevski2020wav2vec,W2V_bert} which have hundreds of thousands of additional training data hours. The more advanced methods such as Self-supervised pre-training are used to utilize those extra hours, so the model can become more resilient to noise, effectively making those systems usable in various scenarios and easily adaptable toward various domains. By the same token, models that use pseudolabelling and additional augmentation methods will not be included. We do not explore performance of fusion approaches introduced in \cite{MFS} and \cite{blstm_fusion}, given that SOTA shows drawbacks in their performance, aleviated in this work. The approach proposed in  \cite{9689650} is an early fusion, not a late fusion method and is thus also omitted. We used the espnet framework \cite{watanabe2018espnet} as the base, and built on top of it; hence our results should be easily reproducible. 

\subsection{Datasets}
We performed evaluation experiments on four publicly available ASR datasets: 1) LibriSpeech \cite{7178964}, containing 1,000 hours of English speech; 2) Aishell-1 \cite{aishell_2017}, featuring 170 hours of Mandarin speech; 3) GigaSpeech \cite{chen21o_interspeech}, with 10,000 hours of English speech; and 4) the Tedlium (v2) \cite{tedliumv2} with 207 hours of Ted talks.

\subsection{Implementation details}
In this work, we used the recent architecture of E-branchformer \cite{kim2022ebranchformer} as a base model in our ensemble, since it has shown the best performance on explored datasets. We used the default espnet \cite{watanabe2018espnet} configurations for the corresponding datasets. The preprocessing generally included the global normalization, spectral augmentation and data perturbations for the robustness. Models were trained between 30 and 80 epochs depending on the default configuration from espnet. The final model was derived by averaging the weights of the top 10 models, selected based on their performance on the validation set, following the same approach used in the ESPnet framework. It was necessary to deviate from the default hyperparameters in certain instances to ensure training stability. We utilized joint CTC-attention decoding with default CTC weight of $0.3$ and $0.7$ for decoder. The weight was scaled in accordance to the number of ensemble models when LM was used. For each additional model used in the ensemble, we increase the weight of the LM by $0.6$. Thus, the total weight of the LM would be $1.2$ if two models are used in the ensemble. Generally, the decoding configurations are default ones used within espnet framework. We took the pre-trained MEL models directly from the authors of espnet \cite{watanabe2018espnet}, as well as the LM with \textit{Librispeech} dataset. Used hyperparameters can be found in the \href{https://docs.google.com/spreadsheets/d/1ySIKVQgq_p5G5TMDKY8ct7iZuiz3Ncd5bsb16obQk1I/edit?gid=0#gid=0}{Excel sheet}.

\section{Ablation}
\label{sec:ablation}
\noindent The ablation experiment studies the impact of additions of individual base models on the overall accuracy of the proposed ensemble. It is performed on the \textit{Librispeech} dataset.  We first evaluate the most accurate model, the E-Branchformer trained using the MEL features, next we add the second most accurate model, the E-Branchformer trained on the MODGD features and evaluate the ensemble of these two models. In the same fashion, we add E-Branchformers trained on GAMMA and CQT features, after which improvements became negligible. 

Table \ref{tab:model_ablation} shows that each added base model improves ensemble accuracy. Thus, we use ensembles with the same base models structure on all studied datasets.

\begin{table}[htbp!]
\caption{Ablation study with comparison of models W/O language model.}
\label{tab:model_ablation}
\centering
\begin{tabular}{lccccccc}
\hline
\textbf{Model}  &    \multicolumn{2}{c}{\textbf{test}} \\
 &   \textbf{clean} & \textbf{other}  \\
\hline
MEL  &  2.14 & 4.55 \\
+ MODGD  & 2.02 & 4.42 \\
\hspace{3mm}+ GAMMA  &  1.98 & 4.28 \\
\hspace{4mm} + CQT  &  1.98 & 4.24 \\
\hline
\end{tabular}
\end{table}

\section{Results}

We compare to approaches that do not utilize pseudolabeling, semi-supervised learning, large language and foundation models. These techniques can be incorporated in our approach.  For decoding the ensemble model, we found the stopping criteria proposed by Watanabe et al. \cite{decoding} to be inefficient for our approach and that this should be explored further. We employed the standard beam search of size $5$ and limited the peak sequence length to $0.6 \times 0.25 \times speech\_length$. This is the downsampled $60\%$ of the encoder output sequence size. 
\begin{table}[htbp!]
\caption{WER(\%) comparison between different models on \textit{Librispeech} dataset.}
\label{tab:performance_comparison_librispeech}
\centering
\begin{tabular}{lccccc}
\hline
 & \textbf{Params}  & \multicolumn{2}{c}{\textbf{Without LM}} & \multicolumn{2}{c}{\textbf{With LM}} \\
 \textbf{Model}  & (M) & \textbf{test} & \textbf{test} & \textbf{test} & \textbf{test} \\
 &  & \textbf{clean} & \textbf{other} & \textbf{clean} & \textbf{other} \\
\hline
\hline
\textbf{Related baselines} & & & & & \\
Conformer (L) \cite{gulati2020conformer} & 118.8  & 2.1 & 4.3 & 1.9 & 3.9 \\
Branchformer \cite{peng2022branchformer} & 116.2  & 2.4 & 5.5 & 2.1 & 4.5 \\
E-Branchformer (L) \cite{kim2022ebranchformer}  & 148.9 & 2.14 & 4.55 & 1.85 & 3.71 \\
Zipformer (L) \cite{yao2024zipformer} & 148.4 & 2.00 & 4.38 & - & - \\
\hline
\textbf{Our work} & & & & & \\
E-Branchformer (MODGD) & 148.92  & 2.08 & 4.73 & 1.86 & 3.87 \\
E-Branchformer (CQT) & 148.92  & 2.33 & 5.45 & 2.06 & 4.46 \\
E-Branchformer (MEL-ASAPP)  & 148.92  & 2.14 & 4.55 & 1.86 & 3.74 \\
E-Branchformer (MFCC) & 148.92  & 2.15 & 4.88 & 1.9 & 3.98 \\
E-Branchformer (SYMLET) & 148.92  & 2.25 & 5.13 & 1.94 & 4.07 \\
E-Branchformer (GAMMA) & 148.92  & 2.2 & 4.87 & 1.86 & 4.06 \\
Ensemble (Late fusion) & 595.68  & \textbf{1.98} & \textbf{4.24} & \textbf{1.79} & \textbf{3.67} \\
\hline
\end{tabular}
\end{table}


\begin{table}[ht!]
\caption{WER(\%) comparison between different models on the \textit{Tedlium\_v2} and \textit{Gigaspeech} dataset and CER(\%)  comparison on the \textit{Aishell-1} dataset.}
\label{tab:performance_comparison_other}
\centering
\begin{tabular}{llccccc}
\hline
 & & \textbf{Params}  & \multicolumn{2}{c}{\textbf{Without LM}}  \\
 \textbf{Dataset} & \textbf{Model}  & (M) & \textbf{dev} & \textbf{test}  \\
\hline
\hline
& \textbf{Related baselines} & & &  \\
& Conformer in ESPnet \cite{watanabe2018espnet} & 35.53 & 7.5 & 7.6 \\
\cline{2-5}
& \textbf{Our work} & & &  \\
& E-Branchformer (CQT) & 35.01  & 8.40  & 8.72  \\
& E-Branchformer (GAMMA) & 35.01  &  7.76 & 7.49  \\
\textit{Tedlium\_v2} & E-Branchformer (MFCC) & 35.01  & 7.66  & 7.71  \\
& E-Branchformer (MODGD) & 35.01  &  7.44 & 7.47  \\
& E-Branchformer (MEL-ASAPP) & 35.01  &  7.29 & 7.10 \\
& E-Branchformer (SYMLET) & 35.01  & 7.78  & 7.50 \\
& Ensemble (Late fusion) & 140.04   & 6.92  & 6.72  \\
& Ensemble (Late fusion)* & 140.04 & \textbf{6.86} & \textbf{6.65} \\
\hline
\hline 
& \textbf{Related baselines} & & &  \\
& Transformer (Chen et.al.) \cite{chen21o_interspeech}  & 87 & 12.30 & 12.30 \\
& E-Branchformer in ESPnet \cite{kim2022ebranchformer} &148.9 &10.6 &10.5\\
& Conformer in WeNet \cite{yao21_interspeech} & 113.2 & 10.7 &10.6 \\
\cline{2-5}
 & \textbf{Our work} & & &  \\
\textit{Gigaspeech}& E-Branchformer (CQT) & 148.92  & 11.54 & 11.65  \\
& E-Branchformer (GAMMA) & 148.92  &  10.62 & 10.75 \\
& E-Branchformer (MFCC) & 148.92  & 10.68 & 10.74 \\
& E-Branchformer (MODGD) & 148.92  & 10.50  & 10.53  \\
& E-Branchformer (MEL-ASAPP) & 148.92  & 10.60& 10.50 \\
& E-Branchformer (SYMLET) & 148.92  & 10.70 & 10.70 \\
& Ensemble (Late fusion) & 595.68   & \textbf{10.22} & \textbf{10.26}  \\
\hline
\hline 
& \textbf{Related baselines} & & &  \\
& Conformer in ESPnet\cite{watanabe2018espnet} & 46.2 & 4.5 & 4.9 \\
& Branchformer \cite{peng2022branchformer} & 45.4  & 4.19 & 4.43 \\
& E-Branchformer  \cite{kim2022ebranchformer}  & 37.88  & 4.2 & 4.5  \\
& Zipformer-L \cite{kim2022ebranchformer}  & 157.3 & 4.03 & 4.28  \\
\cline{2-5}
& \textbf{Our work} & & &  \\
& E-Branchformer (CQT) & 37.88  & 4.71 & 5.15  \\
\textit{Aishell-1}& E-Branchformer (GAMMA) & 37.88  & 4.44 & 4.77 \\
& E-Branchformer (MFCC) & 37.88  & 4.34 & 4.63 \\
& E-Branchformer (MODGD) & 37.88  & 4.24 & 4.54 \\
& E-Branchformer (MEL-ASAPP) & 37.88  & 4.2 & 4.5 \\
& E-Branchformer (SYMLET) & 37.88  & 4.46 & 4.81 \\
& Ensemble (Late fusion) & 151.52    & 3.80 & 4.03 \\
& Ensemble (Late fusion)* & 151.52    & \textbf{3.78} & \textbf{3.99} \\
\hline
\end{tabular}
\end{table}

\noindent The overall results of the proposed ensemble compared to related SOTA approaches, showing $1\%-14\%$ relative improvement over the best performing SOTA approach, can be observed in Tables \ref{tab:performance_comparison_librispeech}, and \ref{tab:performance_comparison_other}. Results marked with an asterisk (*) in Table \ref{tab:performance_comparison_other} indicate the use of a weighted combination based on validation scores, along with a beam size of 10.






\section{Diversity}
\label{sec:diversity}
\begin{figure}
    \centering
    \subfigure[]{\includegraphics[width=0.49\textwidth]{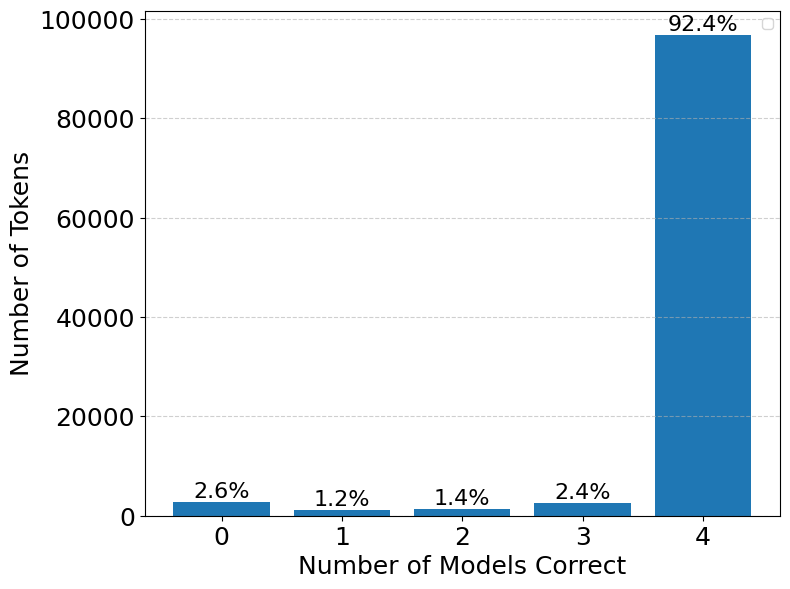}} 
    \subfigure[]{\includegraphics[width=0.49\textwidth]{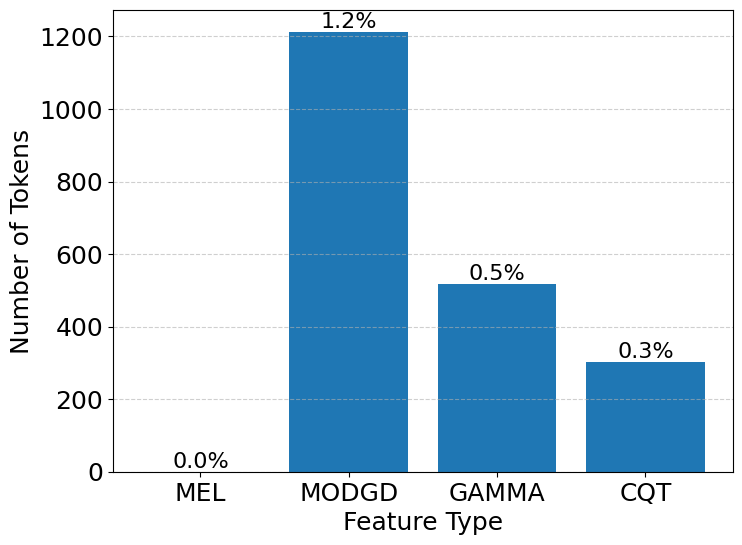}} 
    \caption{(a) Difficulty graph, (b) Incremental theoretical gain}
    \label{Fig:Div}
\end{figure}

As previously stated, the diversity is an important factor that fosters accuracy of an ensemble. We first compute the Difficulty measure \cite{kunchevaETAL} on the \textit{Aishell-1} dataset, by computing the percentage of tokens predicted accurately by all models contained in our ensemble, and combinations of three, two or one model. To make this analysis consistent, we need to ensure that at each step of decoding process, each model is conditioned on the same previous tokens. For this reason we make use of the beam size 1 and apply the token correction at each step, i.e. each following token is predicted conditioned on previous ground truth tokens.
This is presented in Figure \ref{Fig:Div} (a). The \textit{Aishell-1} dataset is the most interesting for these analyses, since each token corresponds to one character. For other datasets, we wouldn't have the equality, rather we would effectively explore the token error rate (TER) whose performance is as well proportional with other metrics of interest. The analyses shows that at least one model from the ensemble makes a mistake on 5\% of the tokens, whereas no model accurately predicts $2.6\%$ of contained tokens. The latter constitutes the current theoretical limit of our ensemble with base models trained using a predefined parameters. Simple strategies such as (weighted) averaging of predictions fall short of this limit. Figure \ref{Fig:Div} (b) studies the percentage of tokens accurately predicted by a model, and not predicted accurately by any of the models preceding it on a chart. The model trained on MEL features is the starting base model, the model trained on MODGD features predicts $1.2\%$ of tokens that were wrongly predicted by the starting model, model trained on GAMMA features predicts $0.5\%$ tokens that were wrongly predicted by both models trained on MEL and on MODGD etc. As the number of models increases, the percentage of additionally accurately predicted tokens decreases. Similar is shown in the ablation study in Section \ref{sec:ablation}.

\noindent The SOTA e-branchformer model trained on MEL features has a CER of $4.57$ within the aforementioned constraints, whereas the proposed ensemble has a CER of $4.07$, which is the improvement of $12.3\%$. Complementarity between models trained on different representations is evident, even for expressive models, such as large e-branchformers. The analyses also reveal possibilities for further improvements in overall accuracy of the proposed ensemble. 

\section{Conclusion}

This work demonstrates that various representations of input speech in combination with SOTA models can still improve performance due to diversity and complementarity of contained information. The proposed ensemble has the advantage of allowing parallel training of base models, that can be of a reduced size. It also allows the flexibility of training different base models on different representations of the data. Using more advanced combination strategies might offer additional improvements in performance of the proposed ensemble.

\begin{credits}
\subsubsection{\ackname} The authors acknowledge the use of supercomputer \texttt{Supek} of the University Computing Center, University of Zagreb. Finally, we wish to express our gratitude to dr. Tomislav Šmuc and Daniil Burakov for valuable discussions.

\end{credits}
%
%
%
 \bibliographystyle{splncs04}
 \bibliography{mybibliography}

\end{document}